\PassOptionsToPackage{xcolor}{dvipsnames}
\documentclass[twocolumn,twocolappendix]{aastex701}
\usepackage{amsmath}
\usepackage{cancel}
\usepackage{wasysym}
\usepackage{xcolor}
\usepackage{tikz}
\usepackage{graphicx}
\usepackage{amssymb}
\usepackage{epstopdf}
\usepackage{mathrsfs}
\usepackage{hyperref}
\usepackage{anyfontsize}
\usepackage{natbib}
\usepackage{color}
\usepackage{lipsum}
\usepackage{diagbox}

\shortauthors{Bandopadhyay et al.}
\shorttitle{Spinning Star rpTDEs}
\begin{document}
\title{The Role of Stellar Spin in Repeating Partial Tidal Disruption Events}
\correspondingauthor{Ananya Bandopadhyay}
\author[0000-0002-5116-844X]{Ananya Bandopadhyay}
\affiliation{Department of Physics, Syracuse University, Syracuse, NY 13210, USA}
\email[show]{abandopa@syr.edu}

\author[0000-0002-9930-3591]{Benjamin Amend}
\affiliation{Department of Physics, Syracuse University, Syracuse, NY 13210, USA}
\email{bjamend@syr.edu}

\author[0000-0003-3765-6401]{Eric R.~Coughlin}
\affiliation{Department of Physics, Syracuse University, Syracuse, NY 13210, USA}
\email{ecoughli@syr.edu}

\author[0000-0002-2137-4146]{C.~J.~Nixon}
\affiliation{School of Physics and Astronomy, Sir William Henry Bragg Building, Woodhouse Ln., University of Leeds, Leeds LS2 9JT, UK}
\email{c.j.nixon@leeds.ac.uk}

\author[0000-0003-1386-7861]{Dheeraj R.~Pasham}
\affiliation{Eureka Scientific, 2452 Delmer Street Suite 100, Oakland, CA 94602-3017, USA}
\affiliation{Department of Physics, The George Washington University, Washington, DC 20052, USA}
\email{p.dheerajreddy@gmail.com}

\author[0000-0002-4043-9400]{T. Wevers}
\affiliation{Astrophysics \& Space Center, Schmidt Sciences, New York, NY 10011, USA}
\affiliation{Space Telescope Science Institute, 3700 San Martin Drive, Baltimore, MD 21218, USA}
\email{twevers@stsci.edu}

\begin{abstract}
The repeated tidal stripping of a star by a supermassive black hole, known as a repeating partial tidal disruption event (rpTDE), can give rise to a transient that rebrightens months to years after the first outburst. Among the rpTDE candidates so far observed, some exhibit dimmer peak luminosities during each successive outburst, which is a trend that has not been reproduced from theoretical models when the star survives more than one encounter with the black hole. Here we suggest that this trend can be recovered if the partially disrupted star is initially (i.e., prior to its first mass-stripping event) rapidly rotating, which is expected if the star was placed on its orbit through the Hills breakup of a tidally locked and tight binary. We test this hypothesis with hydrodynamical simulations of high-mass ($\geqslant 1 M_{\odot}$) main sequence stars repeatedly partially disrupted by a $10^6 M_{\odot}$ black hole, and demonstrate that successively dimmer outbursts are indeed recovered for high (tens of percent breakup) and prograde (i.e., aligned with the orbital angular momentum) stellar spins. Our results provide strong indirect evidence for the operation of the Hills mechanism in seeding the stars in rpTDEs.
\end{abstract}
\section{Introduction}
\label{sec:intro}

A tidal disruption event (TDE) occurs when a star is partially or completely destroyed by the tidal field of a supermassive black hole \citep{hills75,rees88}. Some of the tidally stripped debris remains bound and ultimately accretes onto the black hole, producing an observable flare. The distance at which the tidal field of the black hole becomes comparable to the self-gravity of the star is denoted as the tidal radius $r_{\rm t} = R_\star(M_\bullet/M_\star)^{1/3}$ ($M_\star$ and $R_\star$ are the stellar mass and radius and $M_\bullet$ is the black hole mass).

With the advent of wide-field time-domain surveys, $\gtrsim 100$ TDEs have been detected to date (e.g, \citealp{arcavi14, holoien14, gezari17, pasham18, vanvelzen21, payne21, wevers21, lin22, nicholl22, guolo23, hammerstein23, pasham23, wevers23, yao23,masterson24,guolo24b,yao25b,sfaradi25,ho25}; see also \citealp{gezari21} and references therein). A small subset of these events exhibit rebrightenings on timescales of months to years after the first flare, e.g., ASASSN-14ko \citep{payne21}, AT2018fyk \citep{wevers21,wevers23,pasham24b}, eRASSt-J045650 \citep{liu23,liu24}, AT2022dbl \citep{lin24,hinkle24,makrygianni25}, AT2020vdq \citep{somalwar25}, AT2021aeuk \citep{sun25}, and AT2023uqm \citep{wang25}. These systems, known as repeating partial tidal disruption events (rpTDEs), are speculated to arise from a star orbiting a black hole on a tightly bound orbit with its pericenter near $r_{\rm t}$, with a fraction of its mass stripped on each pericenter passage.

There may be more than one mechanism capable of placing a star on its tightly bound orbit. If the system existed for many orbital periods prior to its first observed outburst, then its present orbital period could reflect the outcome of a combination of gravitational-wave, tidal, and gaseous (e.g., interactions with pre-existing gas in the galactic nucleus) processes that occurred over many orbits \citep{linial24, yao25}. In this scenario, the cumulative effect of tidal heating over multiple orbits could inflate the star, initiating mass transfer once the pericenter distance of the orbit becomes comparable to $\sim \mathrm{few} \times r_{\rm t}$ of the inflated star \citep{yao25}. Secular changes to the eccentricity of a star's orbit arising from gravitational perturbations due to a tertiary perturber (e.g., a companion supermassive black hole) via the eccentric Kozai-Lidov mechanism can also drive the pericenter distance to $\sim \mathrm{few} \times r_{\rm t}$, leading to the partial stripping of the star \citep{melchor24}. However, the eccentric Kozai-Lidov mechanism leaves the semi-major axis of the orbit largely unchanged, thus requiring additional mechanisms to be invoked in order to explain the observed orbital period. Alternatively, the system could be formed in situ via the dynamical capture of a star following the tidal breakup of a tight binary, i.e., Hills capture \citep{hills88,cufari22b,lu23} (it is energetically infeasible to form a sufficiently tight orbit with a partial TDE alone; \citealp{cufari23}).

Analytical and numerical studies have reproduced some properties of rpTDEs. For example, simulations by \citet{bandopadhyay24,liu25} find that low-mass and convective stars lose increasing amounts of mass over multiple pericenter passages as they expand in response to mass loss (consistent with the predictions from \citealp{hjellming87,ge10,zalamea10}) and as they are imparted prograde spin\footnote{\citealp{liu25} argued that the increasing trend in $\dot{M}_{\rm peak}$ is due to the star being strongly thermally inflated via tides, leading to runaway mass loss. However, \citealp{bandopadhyay25} showed that the $1M_\odot$ ZAMS star undergoes an increase in its average density in response to mass loss, and tidal heating has a negligible contribution to the total energy budget, thus the increasing trend in $\Delta M$ and $\dot{M}_{\rm peak}$ is due to the imparted spin, and not tidal heating.} (as the star is spun up, the self-gravity competes with the tidal and centrifugal acceleration terms, leading to an increase in the effective tidal radius and an enhanced susceptibility to mass loss; see Equation 6 of \citealp{golightly19a}). Both of these effects result in progressively brighter flares per outburst, which is consistent with  AT2020vdq \citep{somalwar25} and AT2023uqm \citep{wang25}. On the other hand, mass lost by a high mass ($M_\star \gtrsim 1.5 M_\odot$) and evolved star results in a substantial increase in its average density \citep{hjellming87,dai13,bandopadhyay25}. Such stars can therefore exhibit a declining trend in the amount of mass stripped per orbit. However, the imparted rotation largely counteracts this effect, due to the increase in the effective tidal radius and a resultant shift in the peak timescale to earlier times, thus maintaining a roughly constant peak magnitude of the fallback rate over multiple ($\gtrsim 10$; see Figures 8-11 of \citealt{bandopadhyay24}) orbits. This behavior is consistent with observations of ASASSN-14ko \citep{bandopadhyay24}, but is inconsistent with the progressively \emph{dimmer} outbursts exhibited by AT2018fyk \citealp{wevers23}, eRASSt-J045650 \citealp{liu24}, AT2022dbl \citealp{lin24,hinkle24,makrygianni25}, and AT2021aeuk \citealp{sun25}.

Existing hydrodynamical simulations can produce a dimmer second flare if the star survives only one encounter, i.e., if the majority of its mass is lost on the first encounter and it is completely destroyed during its second pericenter passage (see, e.g., Figure 7 of \citealp{makrygianni25}). This is consistent with the rpTDE candidate AT2022dbl \citep{hinkle26}. The inability to generate progressively dimming flares for more than two orbits presents a notable tension between rpTDE models and observed candidates. Here we suggest that the \emph{initial} spin of the star is the necessary physical ingredient (absent from existing hydrodynamical simulations of rpTDEs, e.g., \citealp{bandopadhyay24,liu25}) that reproduces this effect for more than two outbursts. Specifically, while Hills capture can yield orbits that are consistent with the $\lesssim 1$ year orbital periods of observed rpTDE systems, the required binary separations are sufficiently tight that the binary and stellar binding energies are comparable. If the binary was tidally locked prior to its breakup, the Hills-captured star is then expected to be spinning at a significant fraction of its pericenter angular frequency $\Omega_{\rm p} = \sqrt{(1+e)G M_\bullet/r_{\rm p}^3}$, where $e \simeq 1$ is the orbital eccentricity and $r_{\rm p}$ the pericenter distance. In such instances and with favorable alignments between the initial stellar spin and the angular momentum of the captured-star orbit, the black hole will not efficiently torque the star as it passes through pericenter (as compared to the case in which the star is not initially rotating), plausibly resulting in a dimmer outburst each encounter if the star is of relatively high mass ($M_\star \gtrsim 1M_\odot$, if the mass stripping yields a higher average stellar density).

Here we employ hydrodynamical simulations to show that progressively dimmer outbursts can be actualized from this effect. In Section~\ref{sec:hydro} we show that for high-mass and prograde-spinning stars with angular frequencies comparable to $\Omega_{\rm p}$, and sun-like stars with angular frequencies above $\Omega_{\rm p}$, the tidal interaction with the black hole imparts relatively little additional spin compared to the initially non-spinning case, resulting in dimmer outbursts over multiple encounters. We discuss the implications of our results for observations of rpTDEs, and summarize and conclude in Section~\ref{sec:summary}.

\section{Hydrodynamical Simulations}
\label{sec:hydro}
\subsection{Simulation Setup}
We used the Smoothed Particle Hydrodynamics (SPH) code {\sc phantom} \citep{price18} to model the disruption of main sequence stars with structures derived from {\sc mesa} \citep{paxton11}, each of which is sampled with $10^6$ particles and set uniformly rotating at an angular frequency $\Omega = \lambda \Omega_{\star}$, where $\Omega_{\star} = \sqrt{GM_{\star}/R_{\star}^3}$ and $-1 \lesssim \lambda \lesssim 1$ is the dimensionless spin. We simulate the disruption of stars at zero age main-sequence (ZAMS; when the core hydrogen fraction is $X_{\rm c}\sim 0.7$), middle age main-sequence (MAMS; when $X_{\rm c}\sim 0.2$), and terminal age main-sequence (TAMS; when $X_{\rm c}$ drops below $ \sim 0.001$). Each simulation is initialized with the star on a parabolic orbit about a $10^6 M_\odot$ black hole at a distance of $5 r_{\rm t}$ with pericenter distance $r_{\rm p} = r_{\rm t}/\beta$; see \citet{golightly19a, nixon21, bandopadhyay24} for additional numerical details. Since the near-pericenter dynamics appropriate to a highly eccentric orbit ($e \gtrsim 0.99$, as is relevant for the case of rpTDEs) are effectively equivalent to those of a parabolic orbit, and the fallback rates are likewise quite similar \citep{cufari22a}, these results can be applied to observed systems with a wide range of orbital periods. To model multiple encounters we followed the procedure outlined in \cite{bandopadhyay24} to transport the core back to the initial position of the original star at $\sim2$ days after the star reaches pericenter. By this time the star had settled to a quasi-steady state (as seen in the bottom panels of Figures 1-4). We have verified that our results are effectively independent of the time at which the core is transported (as was also verified in \citealt{bandopadhyay24}). We used the procedures in \citet{miles20,nixon21, bandopadhyay26} to calculate the fallback rates.
\begin{figure}
    \includegraphics[width=0.442\textwidth]{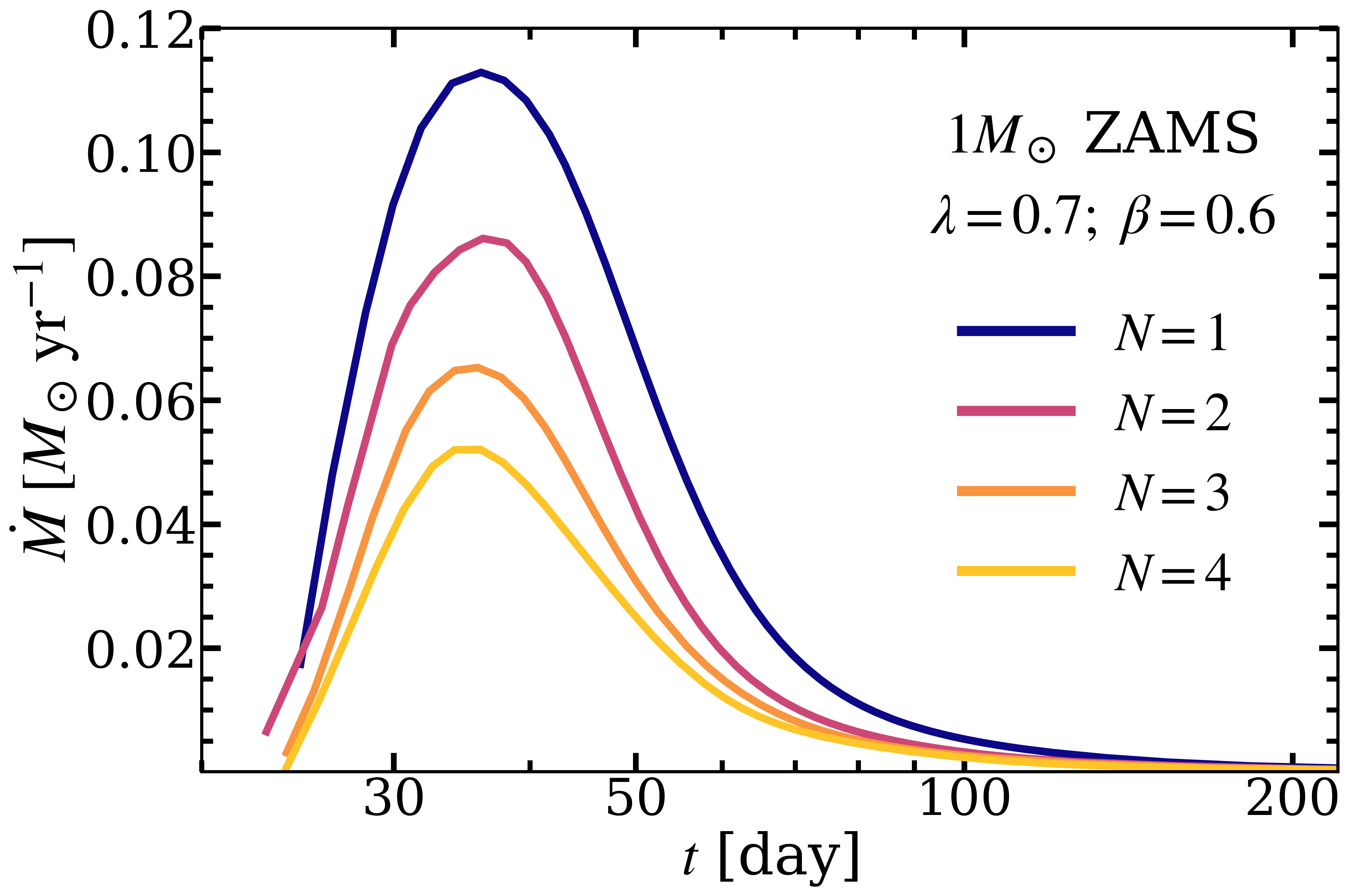} \\
    \includegraphics[width=0.458\textwidth]{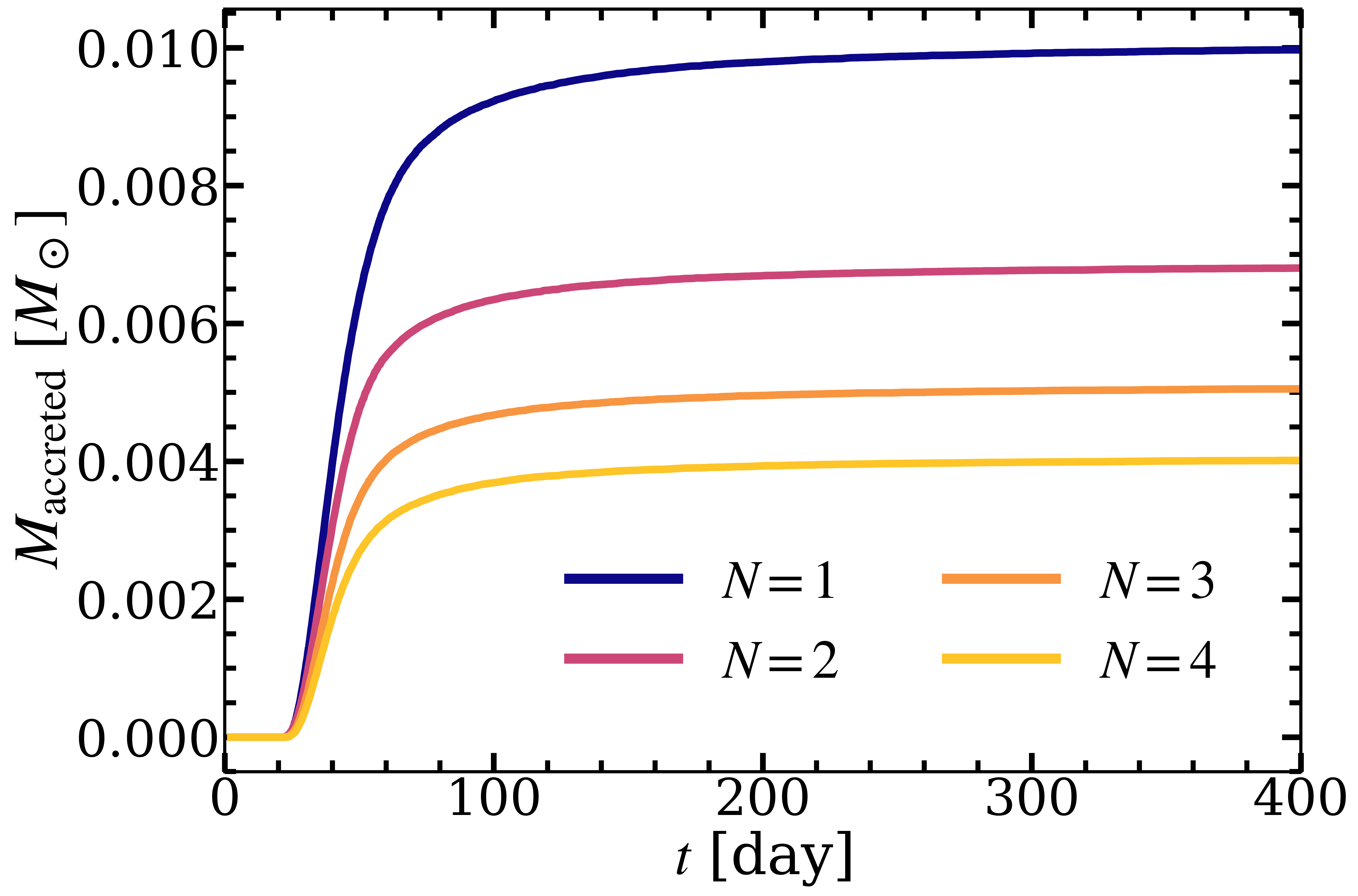} \\
    \includegraphics[width=0.51\textwidth]{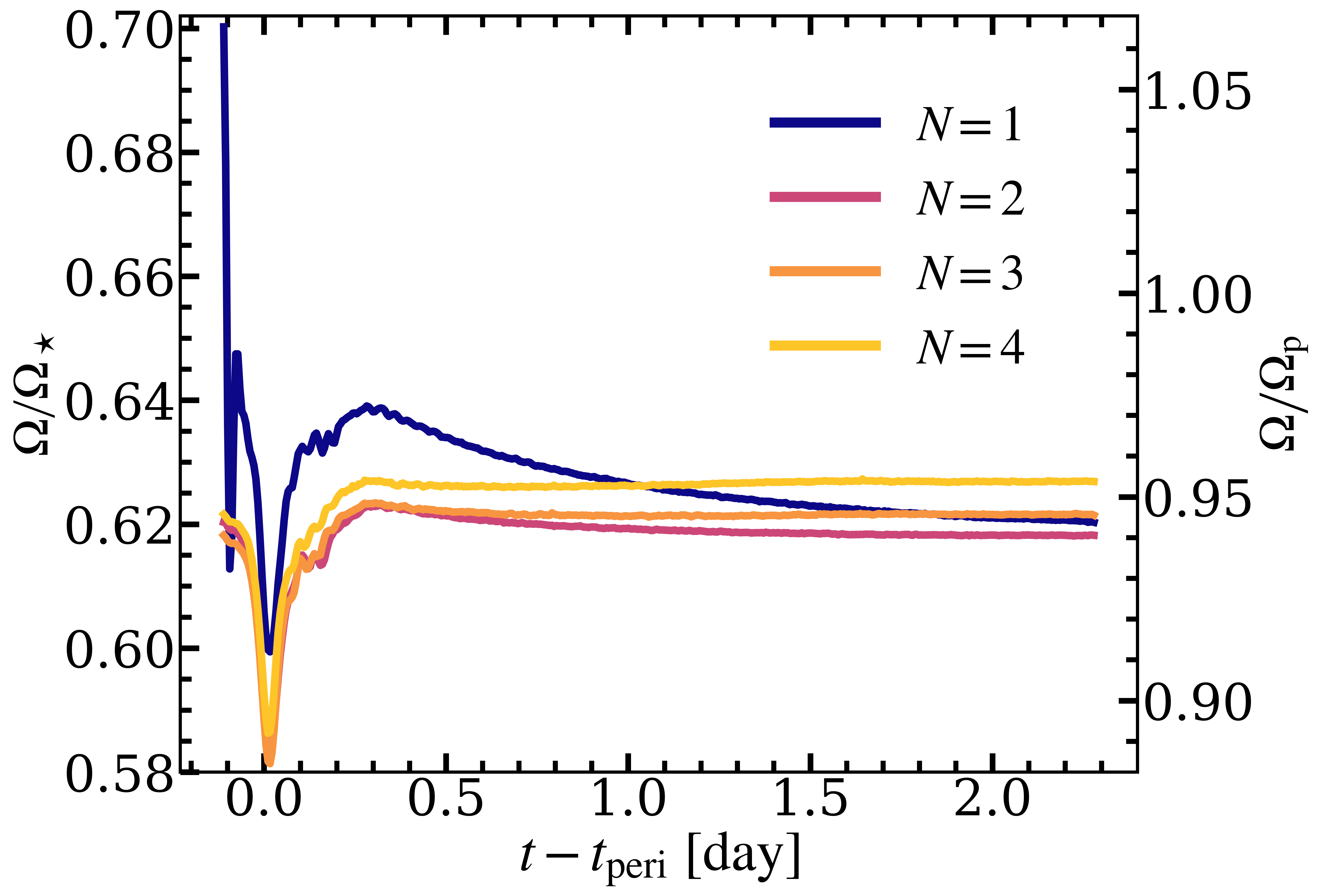}
    \caption{{\bf Top:} Fallback rates for the four pericenter passages of the $1M_\odot$ ZAMS star initially spinning with $\lambda=0.7$, on a $\beta=0.6$ orbit. {\bf Middle:} Mass accreted as a function of time from pericenter passage for the same star and orbit. {\bf Bottom:} Angular velocity of the core (normalized by $\Omega_{\star} = \sqrt{GM_\star/R_\star^3}$ along the primary $y-$axis, and by $\Omega_{\rm p} = \sqrt{(1+e) GM_\bullet/r_{\rm p}^3}$ along the secondary), as a function of time from pericenter passage. With each successive encounter, the amount of mass stripped decreases. Since the spin of the star remains roughly unchanged (relative to the initially non-spinning case), the peak timescale $t_{\rm peak}$ remains roughly constant, and the peak fallback rate ($\dot{M}_{\rm peak} \sim \Delta M / t_{\rm peak}$) exhibits a progressively declining trend.
    } \label{fig:ZAMS1-fallbackrates}
\end{figure}
\subsection{Results}
We simulated the disruption of a $1M_\odot$ ZAMS star with an angular frequency $\lambda=0.7$ (or $\Omega \simeq 1.05 \Omega_{\rm p}$), on a $\beta=0.6$ orbit. The top, middle and bottom panels of Figure~\ref{fig:ZAMS1-fallbackrates} show the fallback rates, the amount of mass accreted (by the black hole), and the angular velocity of the core as a function of time from when the star reaches pericenter for its first four encounters. Since the initial angular velocity $\Omega>\Omega_{\rm p}$, the tidal field torques the stellar spin down to $\Omega \sim 0.62 \Omega_{\star} \approx0.95 \Omega_{\rm p}$ on its first pericenter passage, and it remains effectively unmodified on each subsequent passage.
\begin{figure}
    \includegraphics[width=0.46\textwidth]{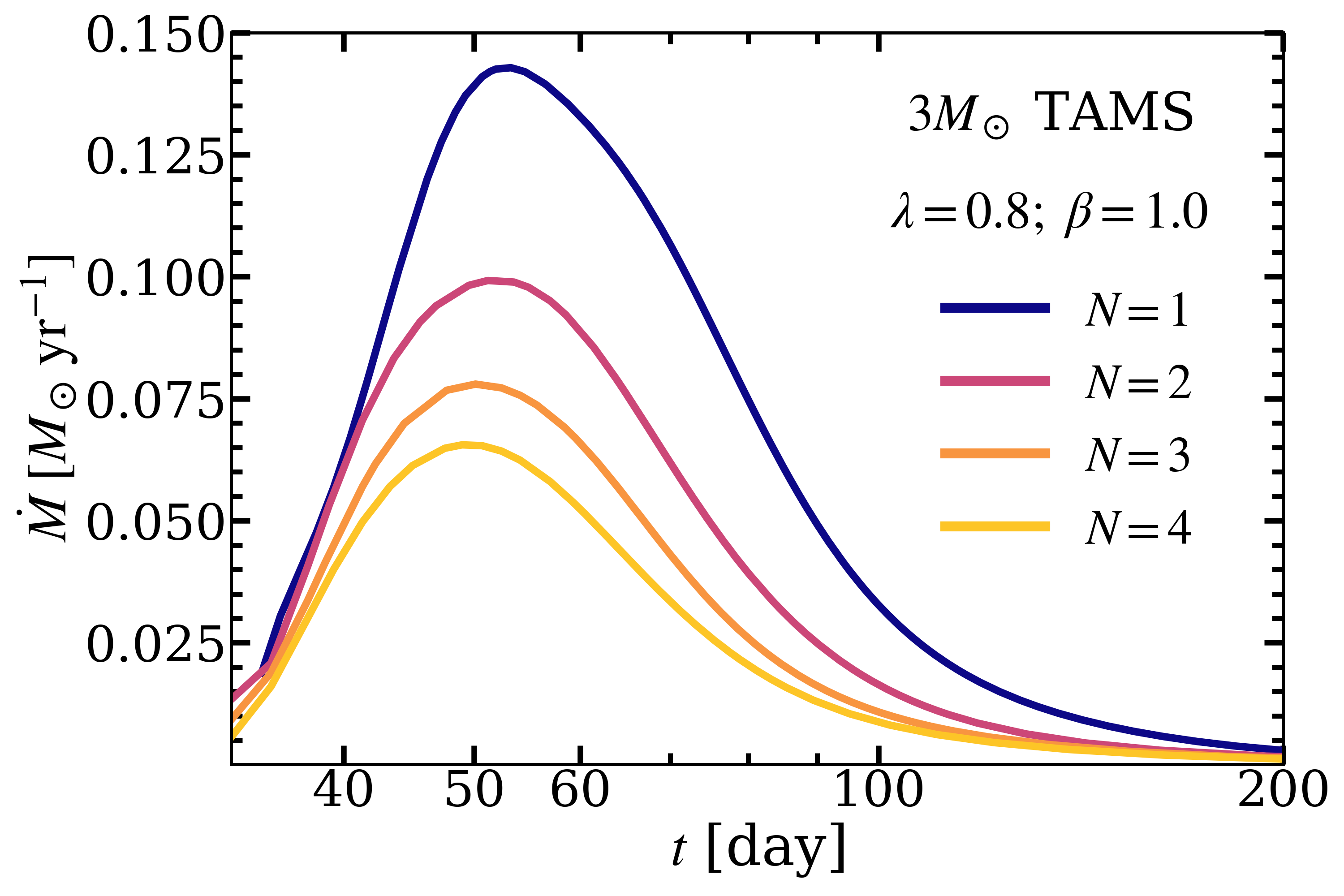} \\
    \includegraphics[width=0.46\textwidth]{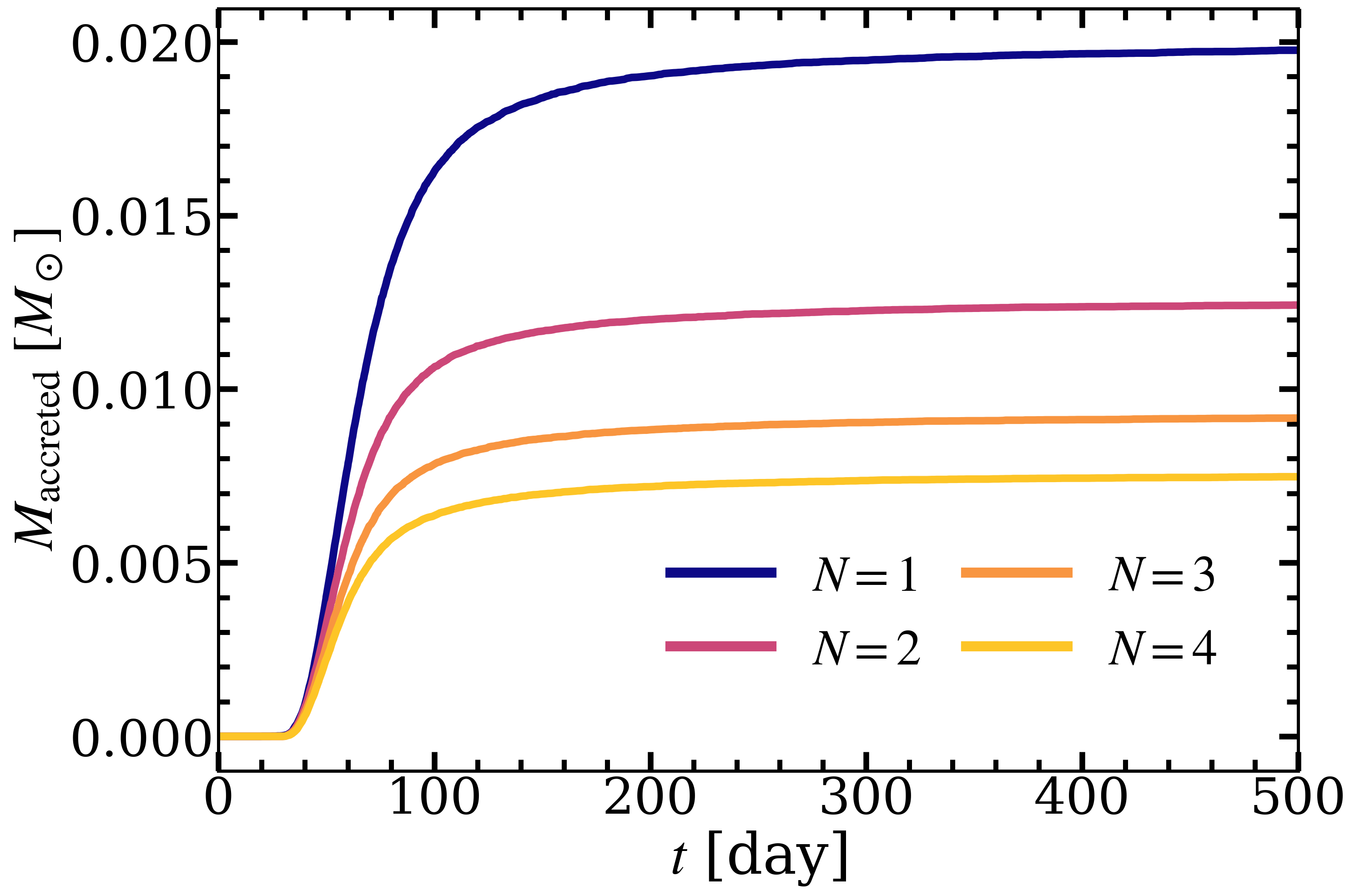} \\
    \includegraphics[width=0.51\textwidth]{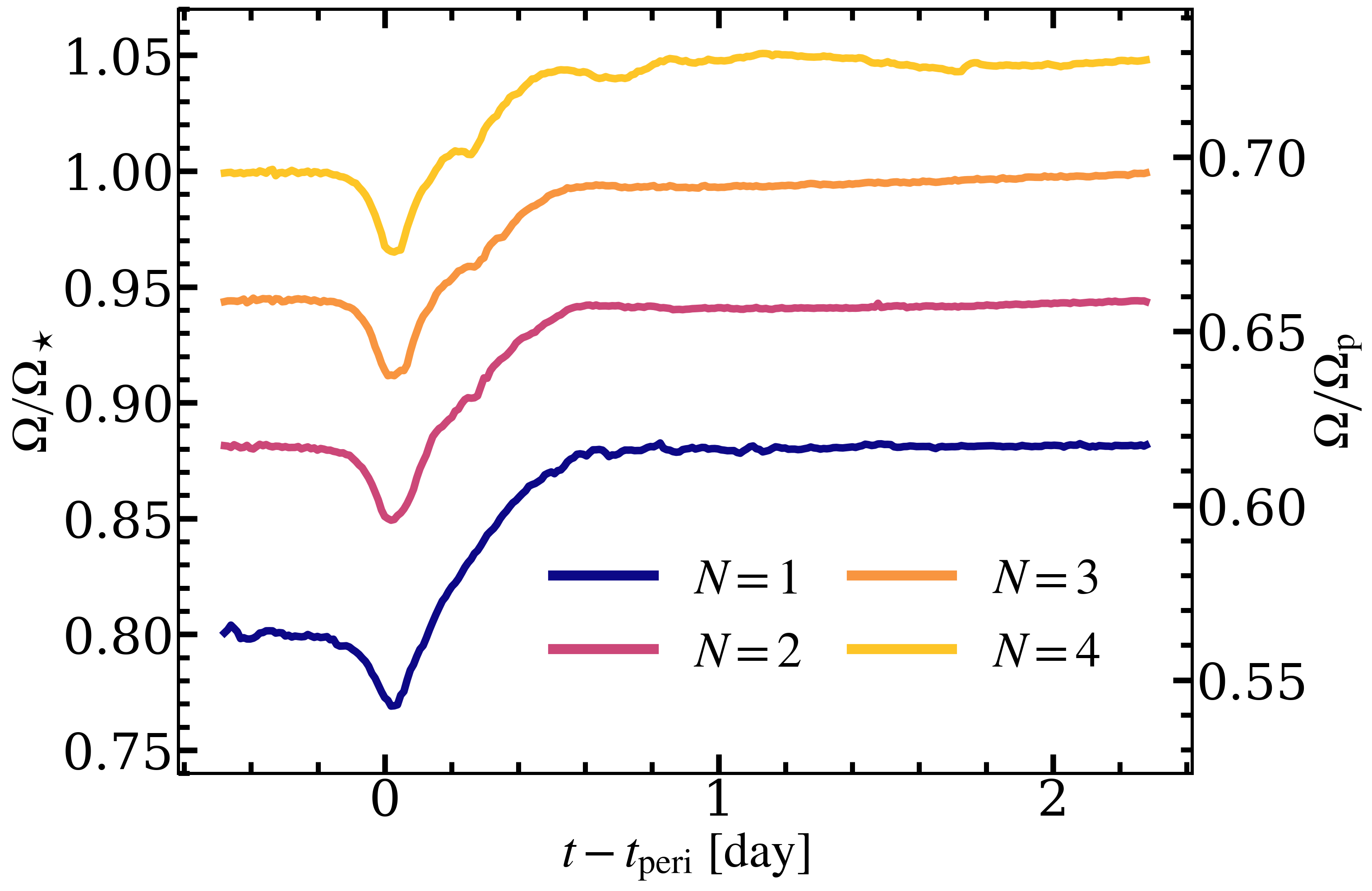}
    \caption{Same as Figure~\ref{fig:ZAMS1-fallbackrates}, but for a $3M_\odot$ TAMS star spinning at $\lambda=0.8$, on a $\beta=1.0$ orbit. Since the initial spin $\Omega$ is a significant fraction of $\Omega_{\rm p}$ the effect of the tidal torque is reduced, rendering the peak timescale $t_{\rm peak}$ relatively unaffected, and leading to a decline in the peak fallback rate.} \label{fig:TAMS3-l0p8-b0p8}
\end{figure}

\begin{figure}
    \includegraphics[width=0.445\textwidth]{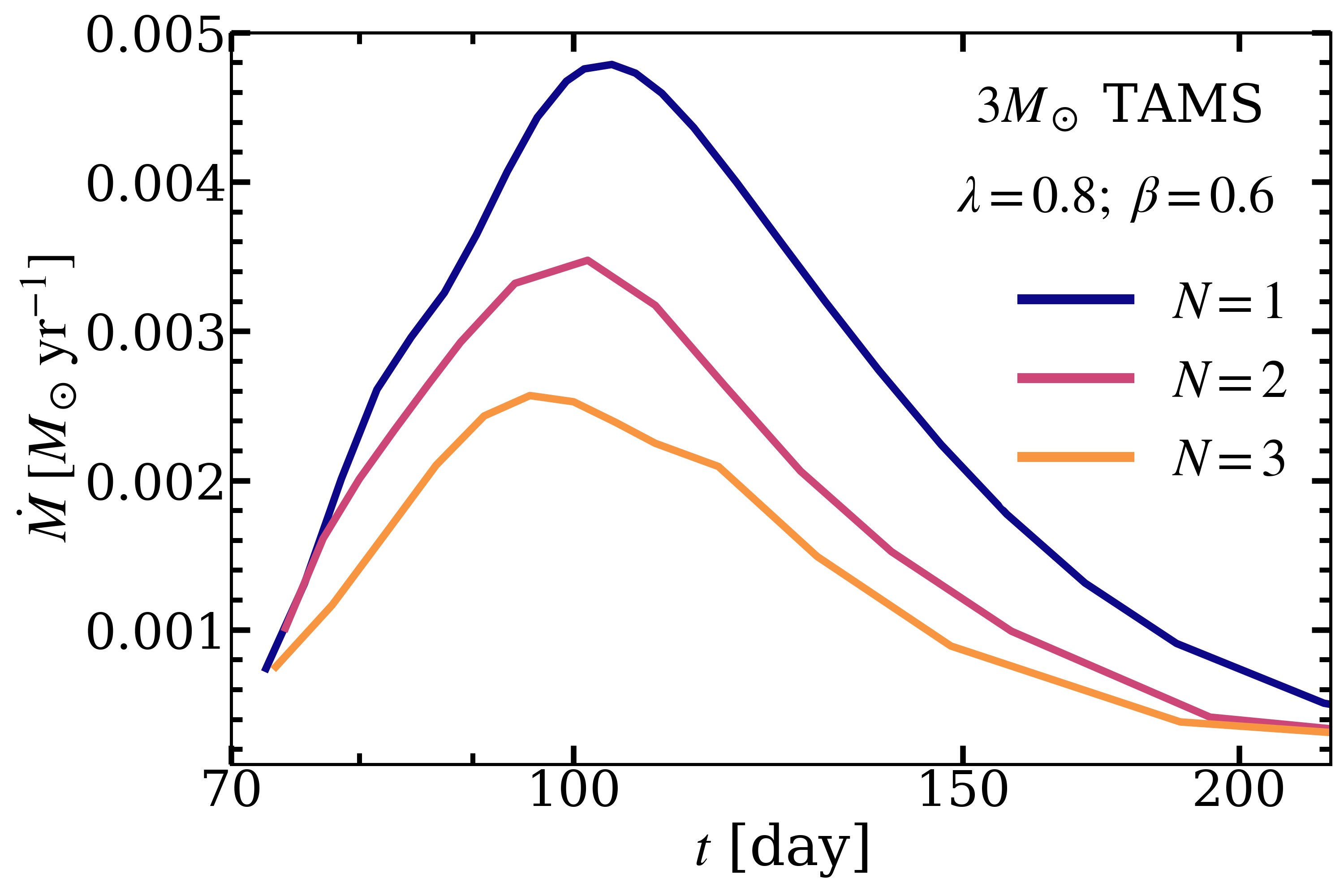} \\
    \includegraphics[width=0.46\textwidth]{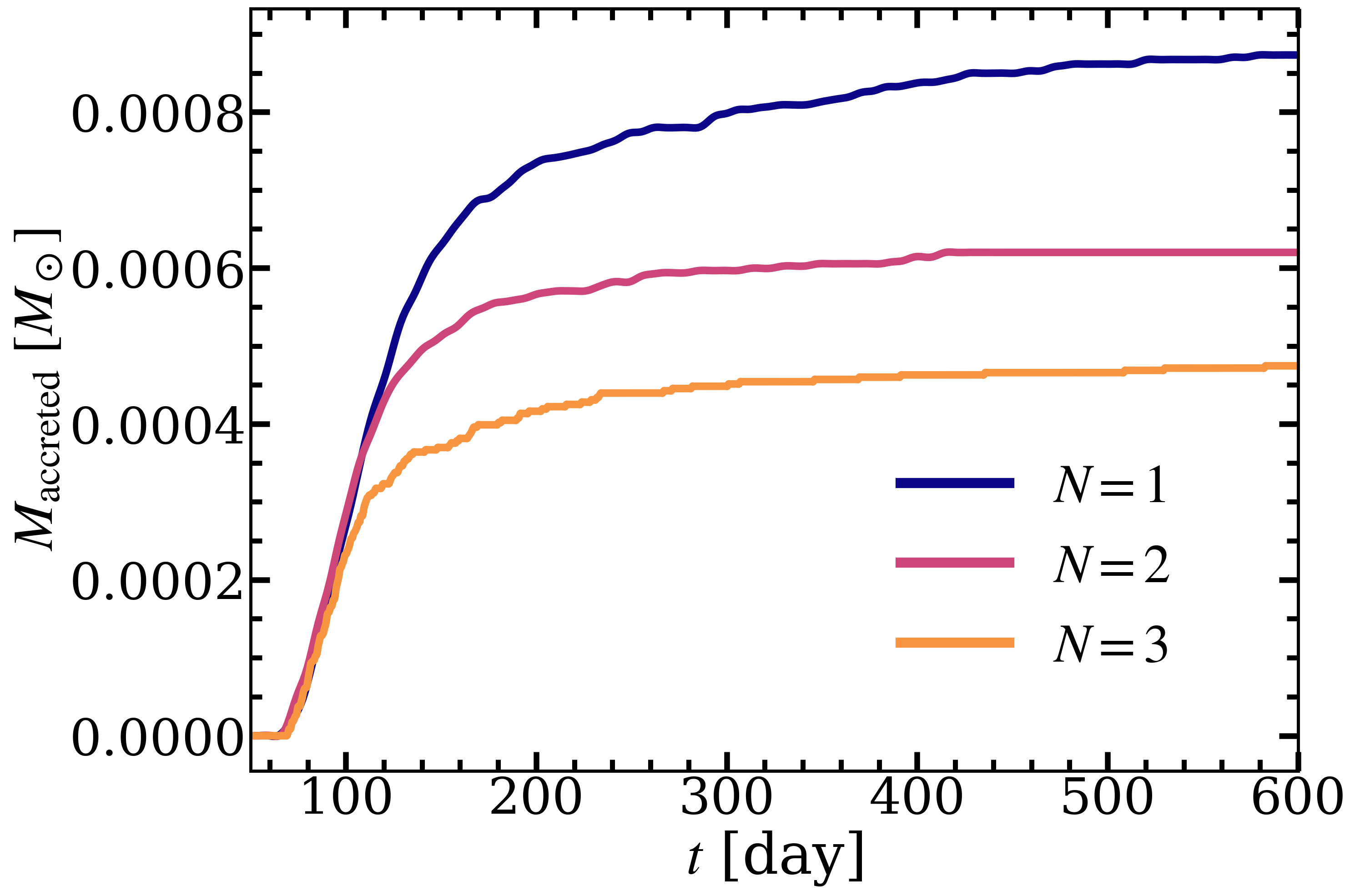} \\
    \includegraphics[width=0.51\textwidth]{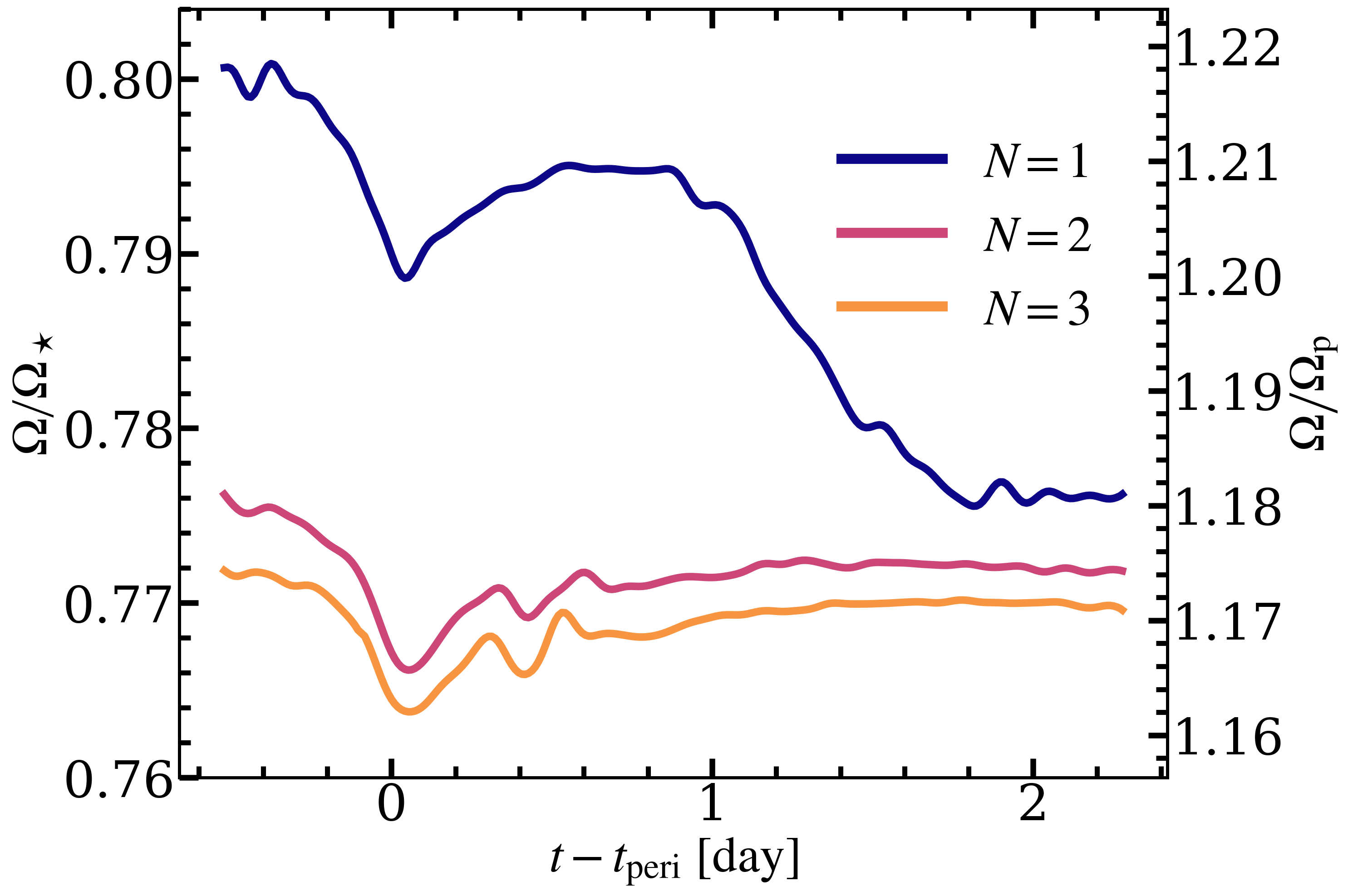}
    \caption{Same as Figure~\ref{fig:ZAMS1-fallbackrates}, but for a $3M_\odot$ TAMS star spinning at $\lambda=0.8$, on a $\beta=0.6$ orbit. Since the initial spin $\Omega > \Omega_{\rm p}$, the tidal torque spins down the star, while the mass stripped and the peak of the fallback rate declines with successive encounters.} \label{fig:TAMS3-l0p8-b0p6}
\end{figure}
\begin{figure}
    \includegraphics[width=0.445\textwidth]{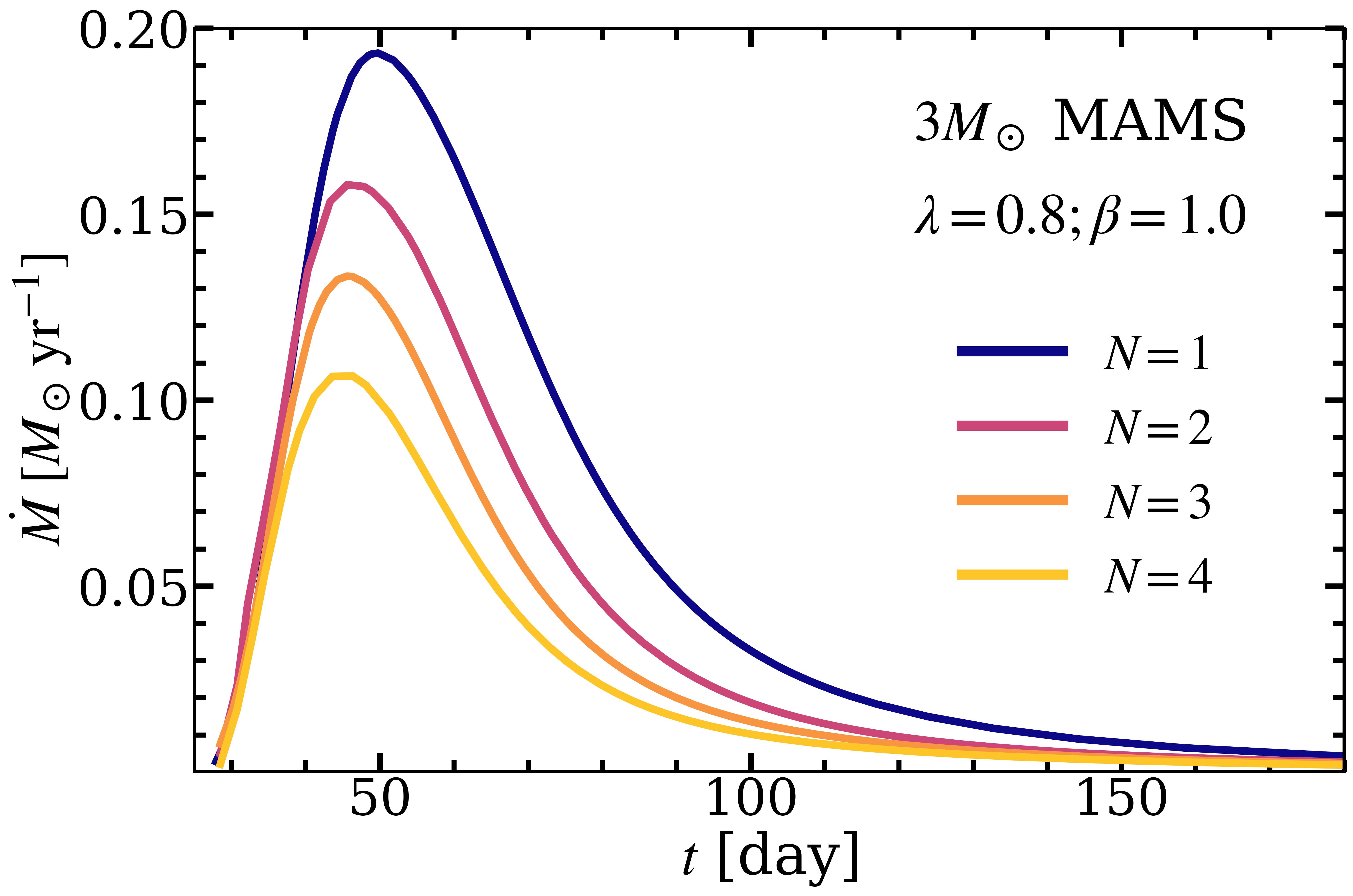} \\
    \includegraphics[width=0.46\textwidth]{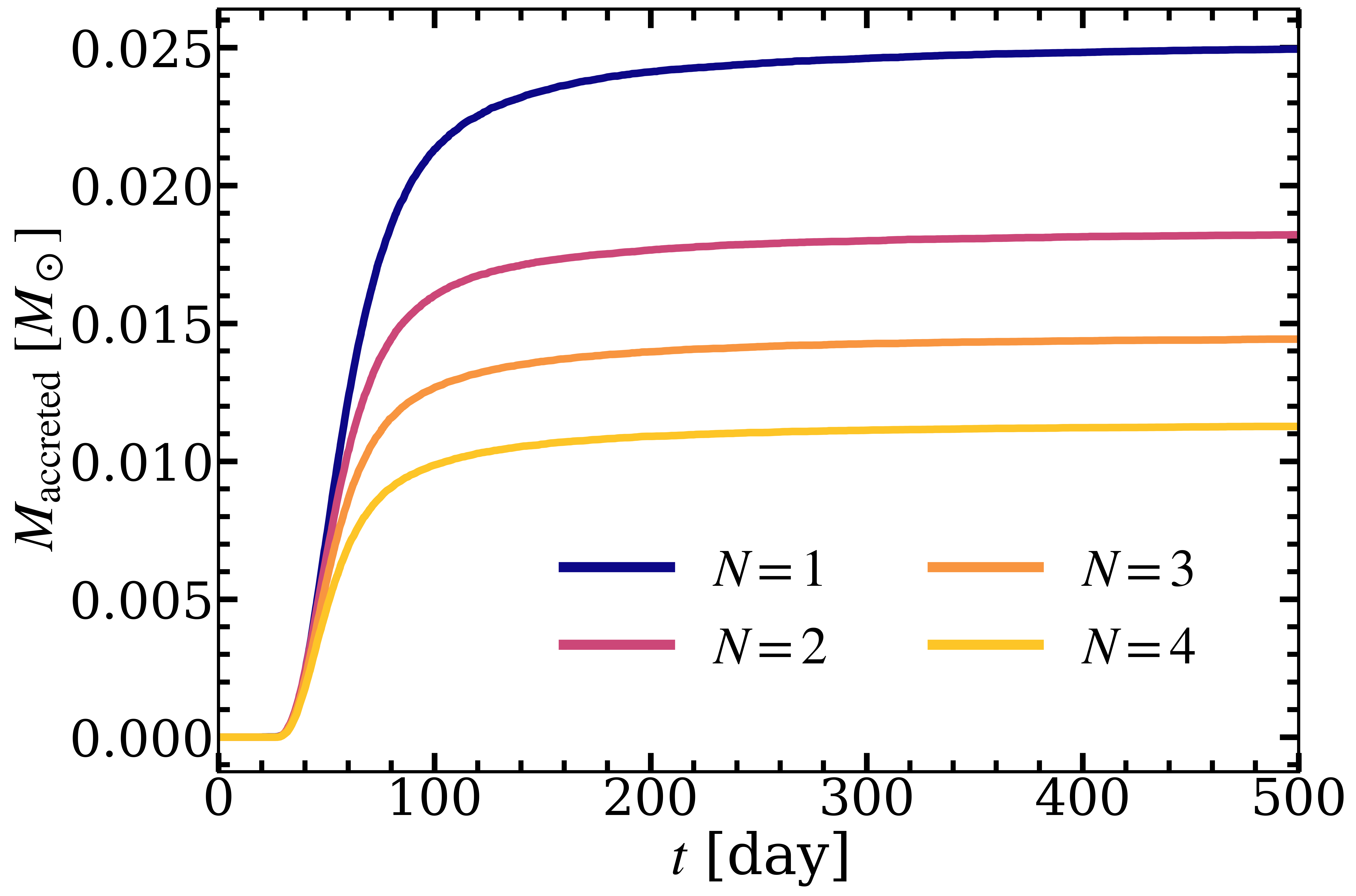} \\
    \includegraphics[width=0.51\textwidth]{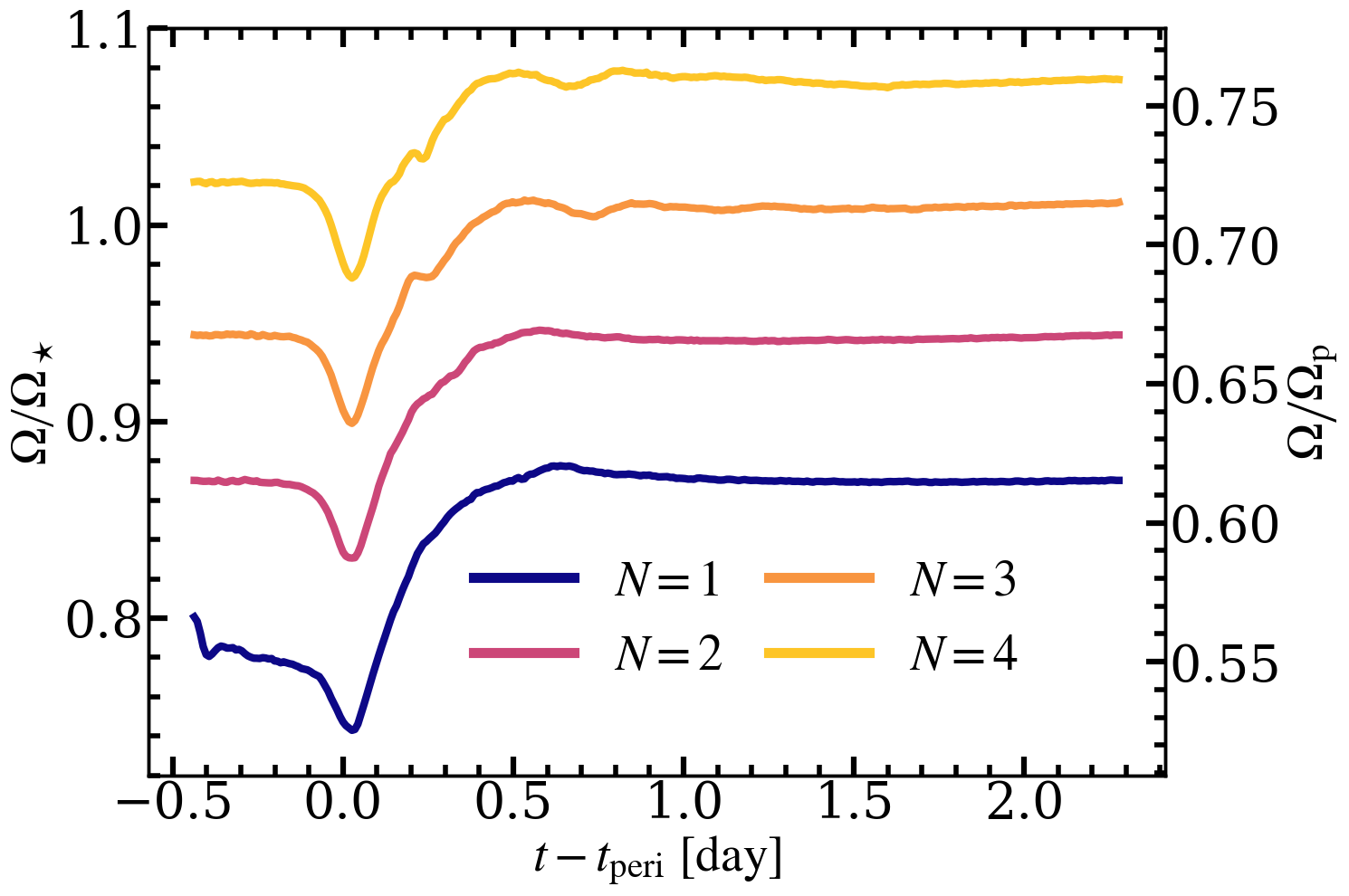}
    \caption{Same as Figure~\ref{fig:ZAMS1-fallbackrates}, but for a $3M_\odot$ MAMS star spinning in a prograde sense relative to the orbital angular momentum, at $\lambda=0.8$, on a $\beta=1.0$ orbit. Since the initial spin $\Omega$ is a significant fraction of $\Omega_{\rm p}$, the effect of the tidal torque is reduced, rendering the peak timescale $t_{\rm peak}$ relatively unaffected, and leading to a decline in the peak fallback rate.} \label{fig:MAMS3-l0p8-b1p0}
\end{figure}

This star undergoes a slight increase in its average density upon the removal of mass. This effect can be understood from an adiabatic mass loss model (\citealp{hjellming87,dai13,bandopadhyay25}; e.g., Figure 5 of \citealp{bandopadhyay25} shows that the average density of this star increases by a factor of $\sim1.2-2$ times its original value for a $1-10\%$ mass loss). Additionally, the tidal interaction spins down the star on its first pericenter passage, while the spin remains roughly constant thereafter at $\sim \Omega_{\rm p}$. The tidal radius of a star spinning at a fraction $\lambda$ of its breakup spin is \citep{golightly19a}
\begin{equation}
    r_{\rm t}(\lambda) = R_\star \left( \frac{M_\bullet}{M_\star}\right)^{1/3} \left( 1\mp \lambda^2 \right)^{-1/3}.
\end{equation}
Thus, for a prograde-spinning star, a reduction in the spin $\lambda = \Omega/\Omega_\star$ corresponds to a decrease in the effective tidal radius and (hence) an increased stability to mass loss. While the spin remains largely unaltered during the subsequent pericenter passages, the star continues to lose decreasing amounts of mass per encounter since its average density increases with each subsequent mass-stripping episode (in accordance with the prediction from the adiabatic mass loss model). As a result, the peak of the fallback rate declines by a factor of $\sim 1.5$ from the first to the second pericenter passage, and by a smaller relative magnitude thereafter. Note that this trend is not reproduced when the star is initially non-spinning, for which the imparted spin generates an increasing $\dot{M}_{\rm peak}$ with each successive outburst (e.g., Figure 13 of \citealt{bandopadhyay24}, see also the simulations in \citealp{kiroglu23,liu25}).

Figure \ref{fig:TAMS3-l0p8-b0p8} shows the results from the partial disruption of a $3M_\odot$ TAMS star with $\lambda=0.8$ and $\beta=1.0$ (or $\Omega \sim 0.55 \Omega_{\rm p}$), which clearly demonstrates a progressive decline in the peak fallback rate (top panel) and a decreasing trend in the amount of mass accreted per encounter (middle panel). The bottom panel of this figure illustrates that the core is spun up slightly on each encounter, which is expected from the fact that $\Omega<\Omega_{\rm p}$ across all four encounters. However, the star can not be spun up beyond $\Omega_{\rm p}$ \citep{bandopadhyay24}, and hence the change in the spin declines with the number of encounters.

Figure \ref{fig:TAMS3-l0p8-b0p6} shows the fallback rates, accreted mass, and angular velocity for the $3M_\odot$ TAMS star spinning with $\lambda=0.8$ and $\beta=0.6$. In this case, the star is spinning at an angular frequency $\Omega>\Omega_{\rm p}$ prior to any mass loss, and the tidal encounter with the black hole spins down the star on each pericenter passage. Additionally, the average density of this star increases in response to mass loss~\citep{bandopadhyay25}, and -- as seen in the middle panel of Figure \ref{fig:TAMS3-l0p8-b0p6} -- loses a progressively declining amount of mass per encounter. The combination of these two effects (reduced spin and reduced mass loss) causes the peak fallback rate to decline, as seen in the top panel of the figure.

Figure~\ref{fig:MAMS3-l0p8-b1p0} shows the fallback rates, accreted mass, and angular velocity for the first four pericenter passages of a $3M_{\odot}$ MAMS star with $\lambda = 0.8$ and $\beta = 1.0$. Similar to the TAMS star, it undergoes a substantial increase in its average density as a result of mass loss. Thus, despite the tidal spin up on this orbit (with $\Omega < \Omega_{\rm p}$), it maintains a declining trend in the amount of mass stripped per encounter and in the peak of the fallback rate.

\section{Summary and conclusions}
\label{sec:summary}
Some rpTDE candidates exhibit multiple flares with successively dimmer peak luminosities, which is a trend that is difficult to reproduce with hydrodynamical simulations over more than two encounters -- if a star loses most of its mass on the first tidal interaction with the black hole, the second outburst can be less luminous than the first, but at the expense of completely destroying it and preventing more than two accretion episodes (e.g., Figure 7 from \citealt{makrygianni25}). While some systems have only displayed two outbursts so far (e.g., AT2022dbl, AT2021aeuk) and could be consistent with the star being completely destroyed on the second encounter, eRASSt-J045650 has thus far exhibited at least 5 outbursts and AT2018fyk showed evidence for a second cutoff in its emission \citep{pasham24b} (implying the survival of the star on its second encounter; \citealt{wevers23}), raising the question of how -- or if -- such systems can be reproduced with the rpTDE model.

Our results in Section~\ref{sec:hydro} show that for relatively high mass stars ($M_\star \gtrsim 1M_\odot$), which undergo an increase in their average density in response to small amounts of mass lost \citep{bandopadhyay25}, if the star already possesses substantial prograde rotation, such that its spin is aligned with its orbital angular momentum and its rotation frequency is comparable to the angular frequency at pericenter $\Omega_{\rm p}$, this successive dimming effect can be reproduced (see Figures \ref{fig:ZAMS1-fallbackrates} -- \ref{fig:MAMS3-l0p8-b1p0}) as a result of the less efficient tidal torque exerted by the black hole. This behavior is consistent with the observed lightcurves of AT2022dbl \citep{hinkle24,makrygianni25}, eRASSt-J045650 \citep{liu23} and AT2021aeuk \citep{sun25}, which exhibit reduced peak luminosities with each outburst by a factor $\lesssim 2$.  As the tidal breakup of a tight binary system is thought to be the means of generating in situ rpTDEs \citep{cufari22b}, the existence of multiple and successively dimmer outbursts provides strong evidence for the operation of this mechanism.

We expect the initial stellar rotation to most efficiently reduce the accretion rate for prograde stellar spins. Conversely, when the spin is retrograde with a magnitude $\lesssim \Omega_{\rm p}$, the relative motion between the star and the black hole (in the co-rotating frame of the star) acts to impart prograde angular momentum to the star. This results in an increased amount of mass lost per encounter and a shorter return time of the debris to the black hole \citep{golightly19a}, the combined effect from which is progressively brighter flares. However, this trend likely does not continue for (retrograde) spins above $\sim \Omega_{\rm p}$, as in this case the black hole makes more than one complete rotation about the star (in the co-rotating frame) near pericenter. Correspondingly the net effect of the tidal torque is likely significantly reduced and the stellar spin remains effectively unmodified; we intend to explore this scenario in future work.

For the rpTDE candidate AT2018fyk, the peak luminosity decreases by approximately an order of magnitude between the first and second outbursts (see Figure~1 of \citealp{wevers23}). To generate a factor of $10$ decline in the peak, our model suggests that the star would likely have to be spinning at an angular velocity $\Omega$ substantially higher than $\Omega_{\rm p}$. For $\beta \gtrsim 0.6-0.8$ orbits (for lower values of $\beta$ high mass stars lose a negligible fraction of their mass, yielding proportionately lower peak luminosities), this would require $\Omega\gg \Omega_\star$, calling into question the stability (and existence) of such systems. However, the inferred black hole mass in AT2018fyk is $M_\bullet \sim 10^{7.7\pm 0.4} M_\odot$, implying that this system possessed a highly relativistic pericenter and that the binary was very tight \citep{wevers23,pasham24}. In this regime, relativistic and chaotic three-body effects could lead to a difference in the pericenter distance of the captured star between the first and second encounter, plausibly generating a larger reduction in the amount of mass stripped and reproducing the observed decline in peak luminosity.

Finally, the Hills mechanism is necessary for generating the short orbital periods in observed rpTDE systems \citep{cufari22b}, and for high-mass black holes -- as is inferred from the $M-\sigma$ relation for some  rpTDE candidates (e.g., ASASSN-14ko, AT2018fyk, eRASSt-J045650) -- the required binary separations are very tight. The tidally locked state of the star then naturally leads to a stellar rotation frequency that is comparable to $\Omega_{\star}$, and the high values of $\lambda$ that generate the declining trend in outburst strength with each successive pericenter passage (Figures \ref{fig:ZAMS1-fallbackrates} -- \ref{fig:MAMS3-l0p8-b1p0}) are manifestly recovered. The Hills mechanism
can therefore place rapidly spinning stars on an orbit around a supermassive black hole that gives rise to progressively dimmer flares, such as those exhibited by eRASSt-J045650. Alternatively, while this leads to a self-consistency constraint on the stellar spin for some observed rpTDE candidates with orbital periods $\sim \mathrm{few} \times100$ days, the star may not be tidally locked with its binary companion and could be rotating at a super-orbital rate. This may alleviate a potential tension that could arise between a wide binary separation required to generate a longer recurrence time (e.g., $\sim 1300$ days for AT2018fyk), and a high value of the stellar spin that would be simultaneously required to generate a substantial drop in luminosity from one outburst to the next. Supporting data for this manuscript is archived on Zenodo \citep{bandopadhyay_2026_20594641}.

\section*{Acknowledgements}
We thank the anonymous referee for useful comments and suggestions that improved the manuscript. A.B.~acknowledges support from NASA through the FINESST program, grant 80NSSC24K1548, and from Syracuse University through the Syracuse University Research Excellence (SURE) Dissertation Grant. B.A. and E.R.C.~acknowledge support from NASA through the Astrophysics Theory Program, grant 80NSSC24K0897, and through Chandra Award Number 25700383 issued by the Chandra X-ray Observatory Center, which is operated by the Smithsonian Astrophysical Observatory for and on behalf of the National Aeronautics and Space Administration under contract NAS8-03060. C.J.N.~acknowledges support from the Leverhulme Trust (grant No.~RPG-2021-380).

\appendix
\begin{figure*}
    \includegraphics[width=0.48\textwidth]{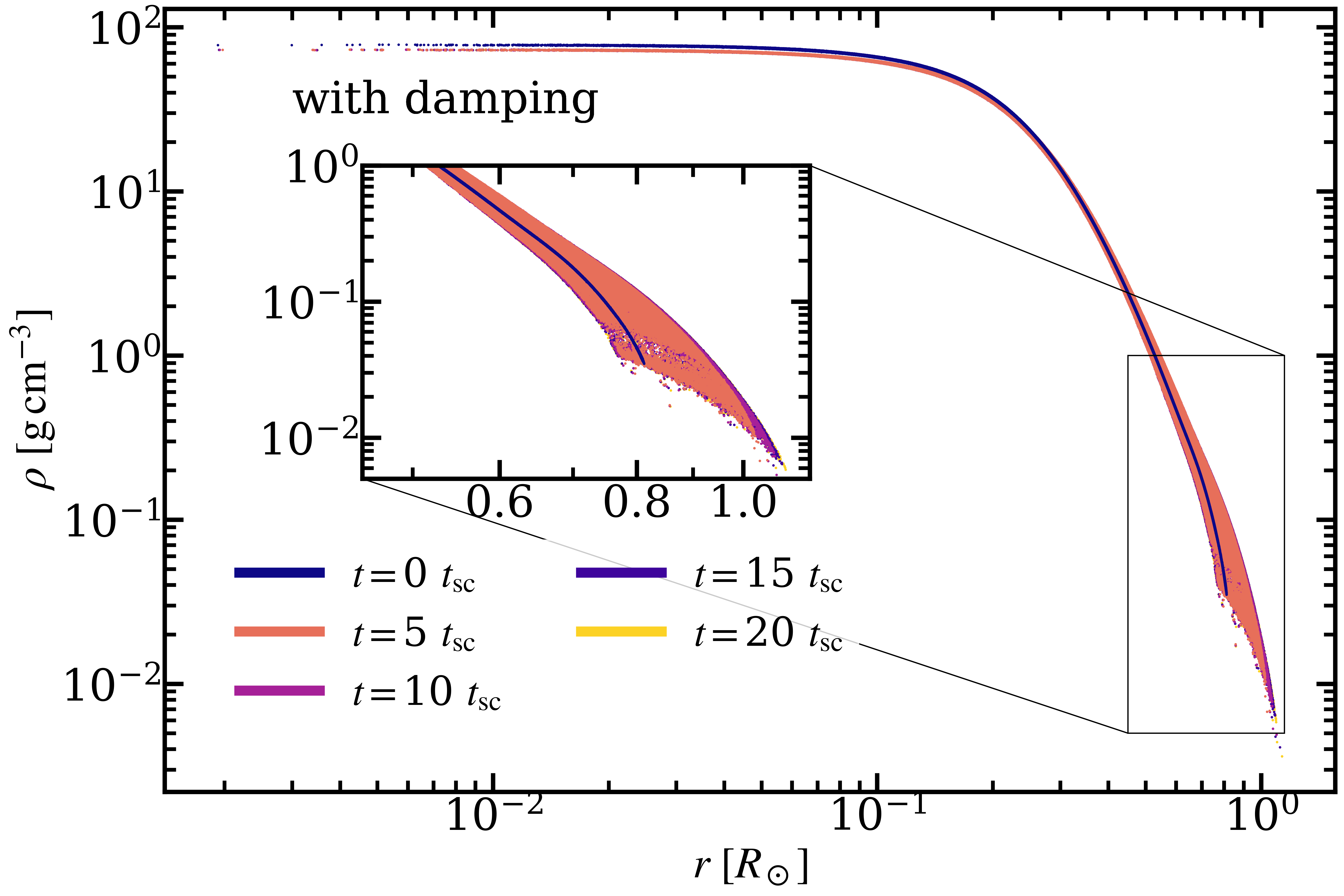}
    \includegraphics[width=0.48\textwidth]{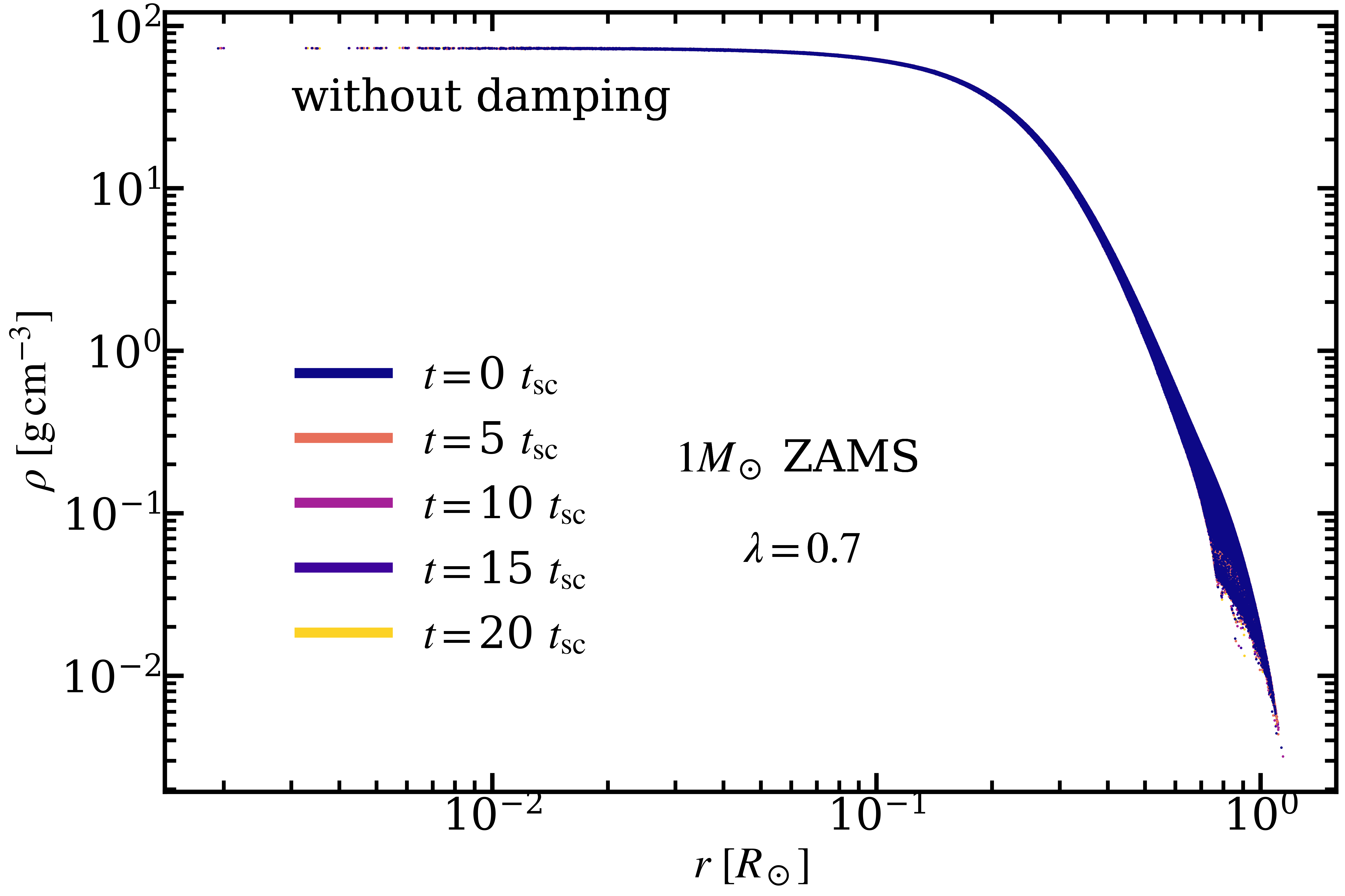} \\
    \includegraphics[width=0.48\textwidth]{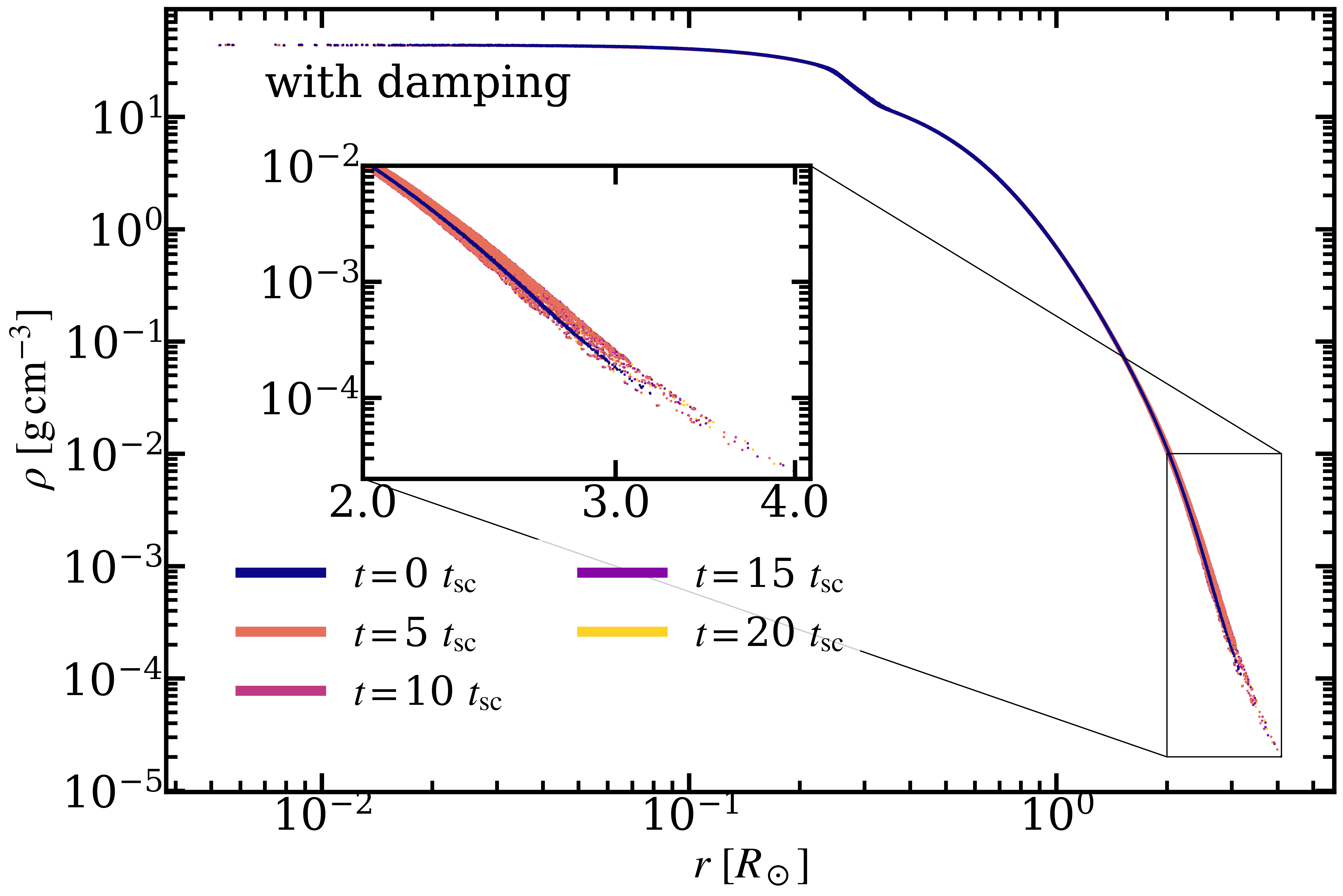}
    \includegraphics[width=0.48\textwidth]{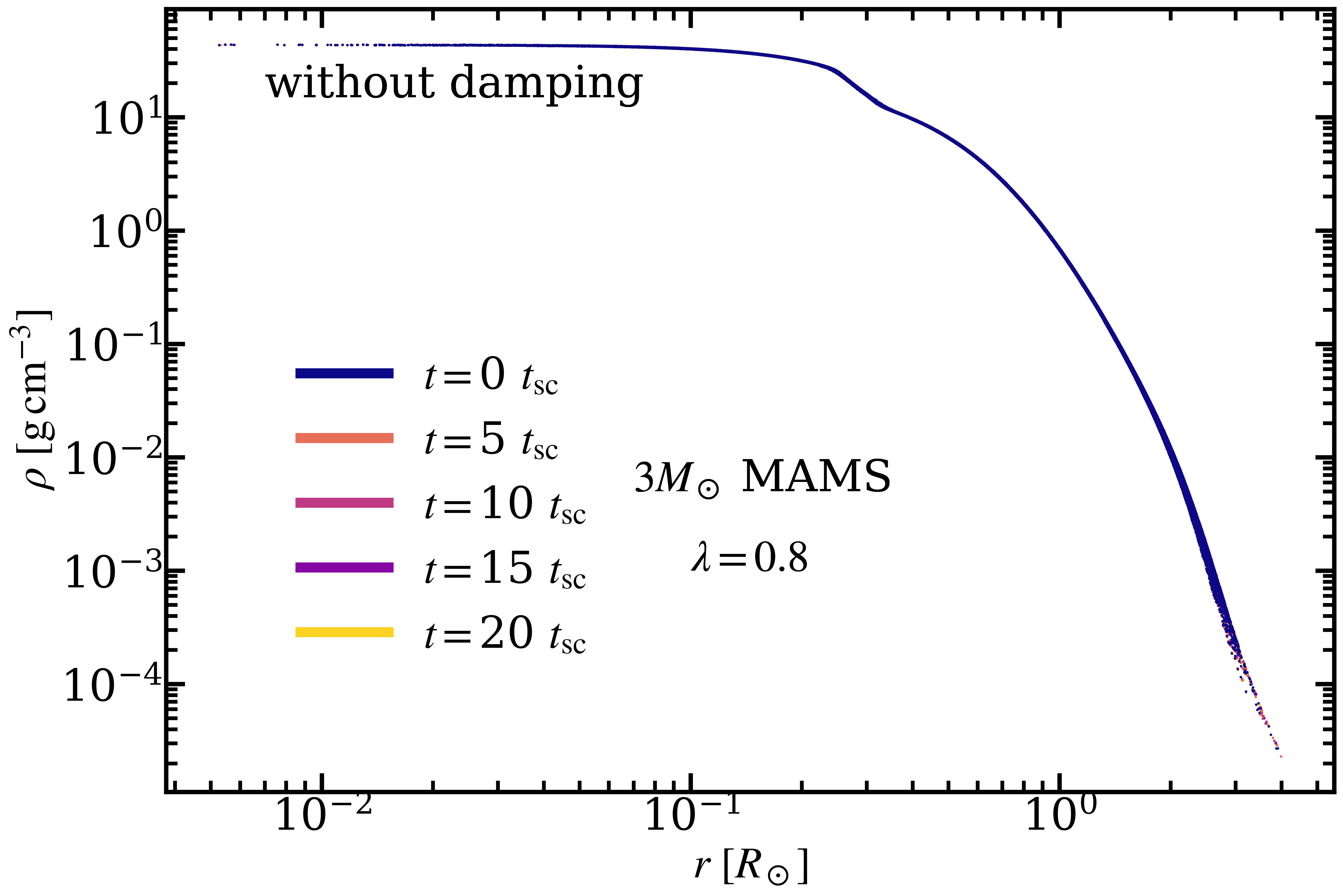} \\
    \includegraphics[width=0.48\textwidth]{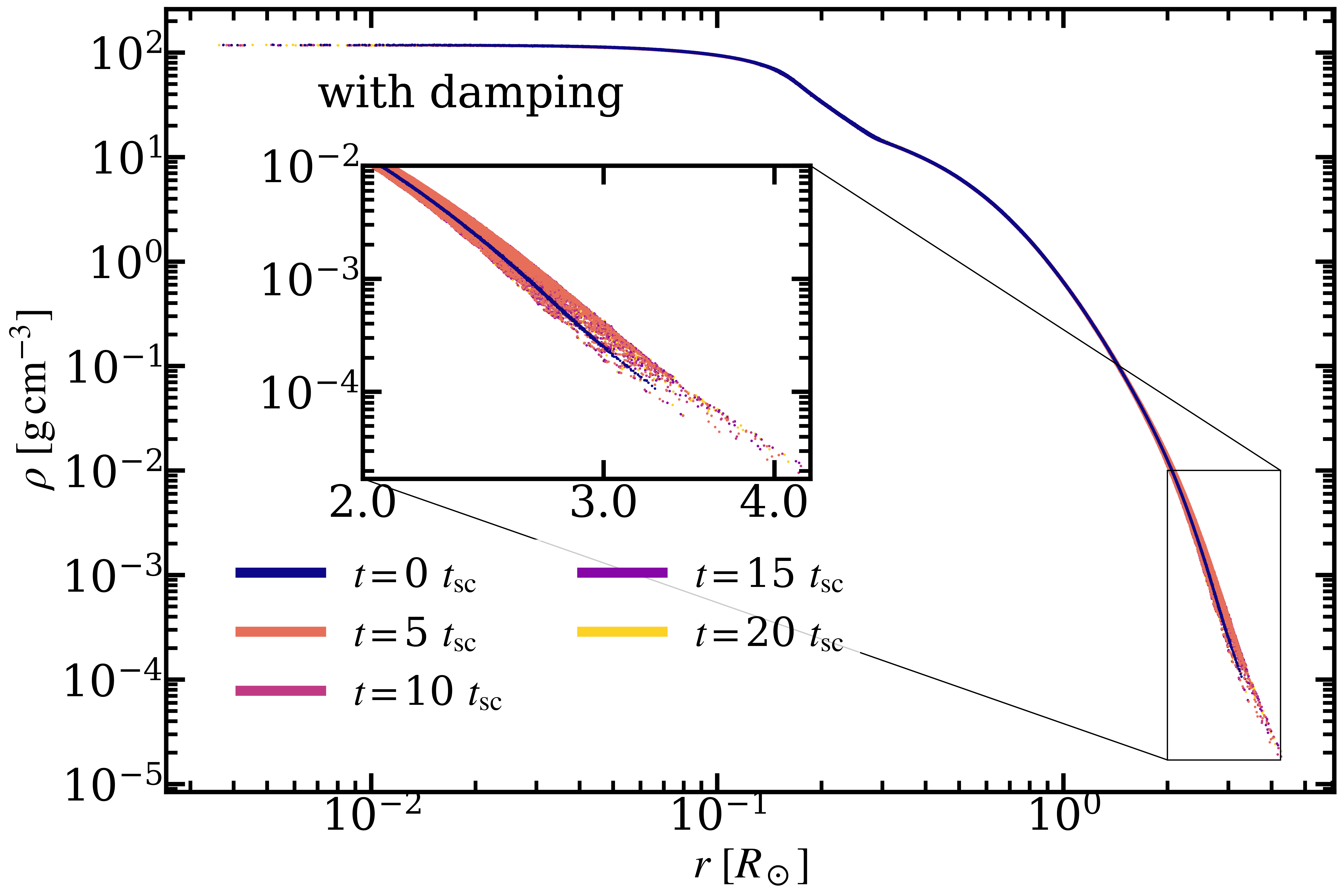}
    \includegraphics[width=0.48\textwidth]{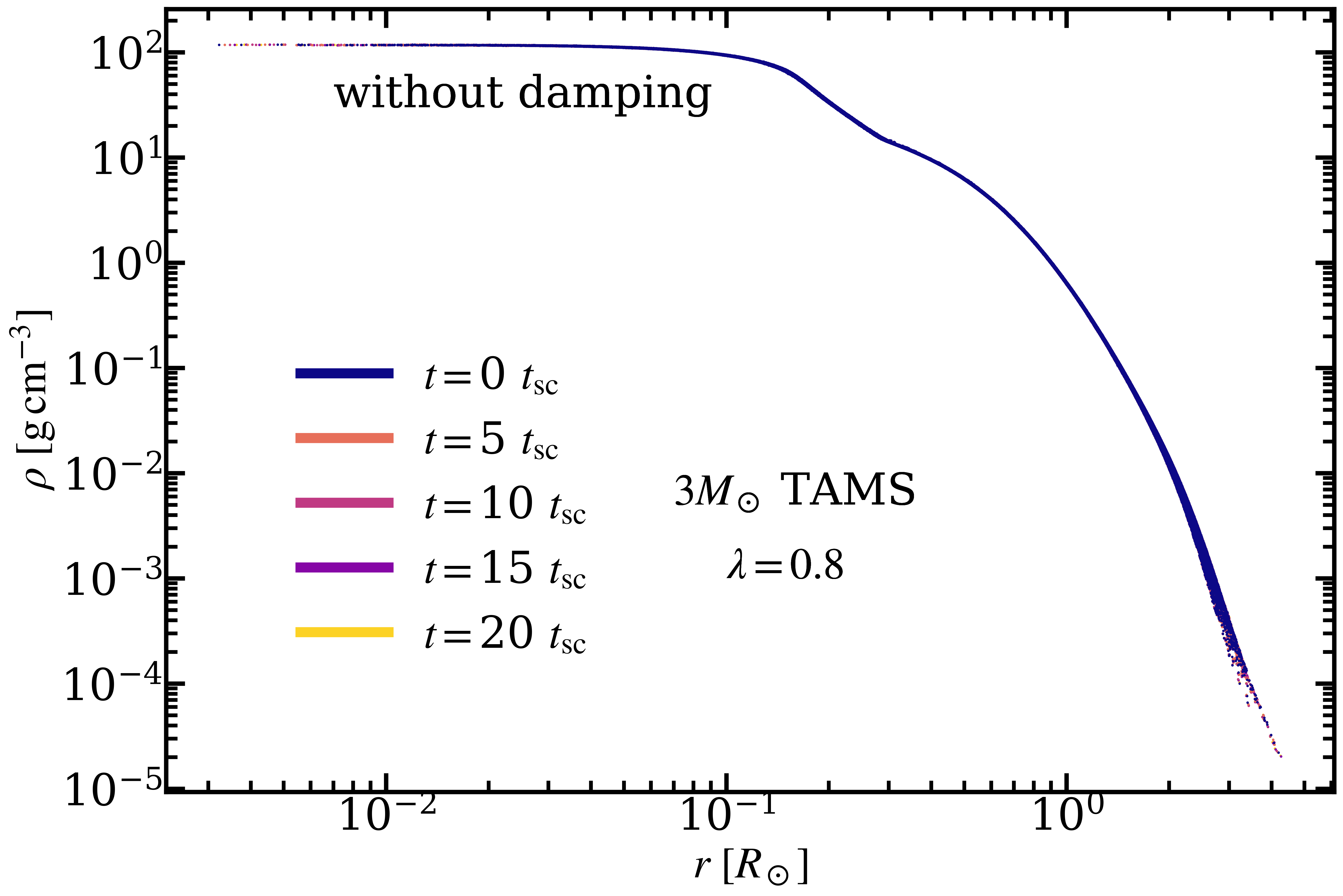} \\
    \caption{Snapshots of the density profile for a $1M_\odot$ ZAMS star with $\lambda=0.7$ (top), a $3M_\odot$ MAMS star with $\lambda=0.8$ (middle panel) a $3M_\odot$ TAMS star with $\lambda= 0.8$ (bottom panel), at different stages in the relax phase with damping (left) and in isolation (right). We note that all of the curves lie essentially on top of one another, demonstrating that our relaxation procedure has produced an accurate, stable configuration.} \label{fig:relaxed-density-profiles}
\end{figure*}
\section{Relaxed Density Profiles in {\sc phantom}}
To set up the spinning stars used for our simulations, we use the approach described in \cite{golightly19b} to relax the density profiles obtained from {\sc mesa} with a velocity damping force for $10$ sound crossing times across the stellar radius, following which we relax them in a co-rotating potential with a constant co-rotational frequency $\Omega_{\rm cor} = \lambda \Omega_\star$, for $20$ sound crossing times. Subsequently, we evolve them in isolation (without the velocity damping force) for another $20$ sound crossing times, to verify the stability of the numerically relaxed configuration. The top panel of Figure~\ref{fig:relaxed-density-profiles} shows snapshots from the relax phase (with damping) and the isolated evolution (without damping) for the $1M_\odot$ ZAMS star with $\Omega_{\rm cor} = 0.7 \Omega_\star$. During the damping phase, the star expands in response to the centrifugal potential, and readjusts to a new equilibrium. The isolated evolution phase shown in the right panel of the figure demonstrates that the equilibrium is stable, and the star maintains its structure across this phase. The middle and bottom panels of the figure depicts snapshots of the density profile for the $3M_\odot$ MAMS star with $\lambda=0.8$ and the $3M_\odot$ TAMS star with $\lambda=0.8$, which exhibit a similar behavior.

\end{document}